\documentclass[
    ,final            
,epsf,aps
  ]
  {aipproc}

\usepackage{rotating}
\usepackage{graphicx}
\usepackage{pstricks,pst-node}
\usepackage{fancybox}

\newrgbcolor{lightgreen}{.5 1 .5}
\newrgbcolor{lightred}{1 .3 .3}
\newrgbcolor{lightgray}{.8 .8 .8}
\newrgbcolor{gray}{.5 .5 .5}
\newrgbcolor{darkgray}{.2 .2 .2}

\layoutstyle{8x11double}

\def\lsim{\ \rlap{\raise 3pt \hbox{$<$}}{\lower 3pt \hbox{$\sim$}}\ }
\def\gsim{\ \rlap{\raise 3pt \hbox{$>$}}{\lower 3pt \hbox{$\sim$}}\ }

\def\ncr{\langle r_V^2 \rangle}
\def\acr{\langle r_A^2 \rangle}
\def\ncrt{\langle r_V^2(\nu_\tau)  \rangle}
\def\acrt{\langle r_A^2(\nu_\tau) \rangle}
\def\vacrt{r_{V+A}^2(\nu_\tau) }
\def\ncrm{\langle r_V^2(\nu_\mu) \rangle}
\def\acrm{\langle r_A^2(\nu_\mu) \rangle}
\def\vacrm{r_{V+A}^2(\nu_\mu)}
\def\swsq{{\sin^2\theta_W}}

\begin{document}

\title{  
On the neutrino vector and axial vector charge
radius\footnote{Plenary talk given in the X Mexican School of
Particles and Fields, Playa del Carmen, México.
Based in part on {\tt hep-ph/0210137} written in
collaboration with M. Hirsch and D. A. Restrepo.}}

\author{Enrico Nardi}{
  address={INFN -- Laboratori Nazionali di Frascati, C.P. 13, I00044 Frascati,
  Italy  \\
{\rm and} \\ 
Instituto de F{\'\i}sica,  Universidad de Antioquia  A.A. {\it
      1226}  Medell{\'\i}n, Colombia  }
}

\begin{abstract}
  A Majorana neutrino is characterized by just one flavor diagonal
  electromagnetic form factor: the anapole moment, that in the static
  limit corresponds to the axial vector charge radius $\left<
  r^2_A\right>$.  As is the case for the vector charge radius of a
  Dirac neutrino, proving that this quantity is a well defined
  physical quantity is non trivial. I will first describe briefly the
  origin of the long standing controversy about the physical or non
  physical nature of the neutrino charge radius. Then I will argue
  that, in contrast to Dirac neutrino electromagnetic form factors,
  for Majorana neutrinos cosmological and astrophysical arguments do
  not provide useful informations on $\left< r^2_A\right>$. Therefore
  this quantity has to be studied by means of terrestrial
  experiment. Finally, I will discuss the constraints that can be
  derived on $\left< r^2_A\right>$  for the tau neutrino from a
  comprehensive analysis of the data on single photon production off
  $Z$-resonance, and I will conclude with a few comments on $\nu_\mu$
  scattering data from the NuTeV, E734, CCFR and CHARM-II
  collaborations and on the limits implied for $\left<
  r^2_A\right>$ for the muon neutrino.
\end{abstract}

\maketitle


\section{Introduction}

Experimental evidences for neutrino oscillations
\cite{SK,Ahmad:2002jz} imply that neutrinos are the first elementary
particles whose properties cannot be fully described within the
Standard Model (SM).  This hints to the possibility that other
neutrino properties might substantially deviate from the SM
predictions, and is presently motivating vigorous efforts, both on the
theoretical and experimental sides, to understand more in depth the
physics of neutrinos and of their interactions.  

Neutrinos electromagnetic interactions can play an important role in a
wide variety of domains, as for example in cosmology
\cite{Dolgov:2002wy} and in astrophysics
\cite{Mohapatra:rq,Raffelt:wa}.  The electromagnetic properties of
Dirac neutrinos are described in terms of four form factors. The
matrix element of the electromagnetic current between an initial
neutrino state $\nu_i$ with momentum $p_i$ and a final state $\nu_j$
with momentum $p_j$ reads \cite{Nieves:1981zt,Shrock:1982sc}
\begin{eqnarray}
\label{JmuDirac}
\langle \nu^D_j(p_j)\,|\, J_\mu^{\rm EM} \,|\, \nu^D_i(p_i)
\rangle\!\! 
&=&\!\! i\bar{u}_j \Gamma_\mu^D(q^2) u_i\,; \\ 
 \Gamma_\mu^D(q^2) = (q^2 \gamma_\mu-q_\mu q\!\!\!\!/) && \hspace{-.9truecm}
[V^D(q^2) - A^D(q^2) \gamma_5]  + \nonumber  \\  
&& \hspace{-2truecm}
i \sigma_{\mu\nu}q^\nu [M^D(q^2)+E^D(q^2)\gamma_5]\,,  
\end{eqnarray}
where $q=p_j-p_i$, and the $(ji)$ indexes denoting the relevant elements of
the form factor matrices have been left implicit.  In the $j=i$ diagonal case,
$M^D$ and $E^D$ are called the magnetic and electric form factors, that in the
limit $q^2\to 0$ define respectively the neutrino magnetic moment $\mu=M^D(0)$ and
the (CP violating) electric dipole moment $\epsilon=E^D(0)$, while the  form
factors $q^2 V^D(q^2)$ and $q^2 A^D(q^2)$ correspond to non vanishing neutrino
charge distributions of the form 
\begin{equation} 
\rho_V(r)=\int\frac{d^3q}{(2\pi)^3}\>  q^2V(q^2)\> e^{i\vec q\cdot\vec r}
 \nonumber
\end{equation}
that are induced by the virtual transitions $\nu_l \rightleftharpoons
l^- W^+$ ($l=e, \mu, \tau $).  Intuitively, these distributions can be
depicted as a positive core surrounded by a negative cloud:

\bigskip
\bigskip
\bigskip

\rnode{na}{}\hspace{1cm}\rnode{naa}{}\hspace{2mm}\rnode{naaa}{}\hspace{.9cm}\rnode{nb}{}%
\hspace{2.4cm}
\rnode{nc}{}\hspace{1cm}\rnode{ncc}{}\hspace{2mm}\rnode{nccc}{}\hspace{.9cm}\rnode{nd}{}
\ncline[linewidth=3pt,linecolor=black]{->}{na}{naaa}\Aput{$\mathbf\nu$}
\ncline[linewidth=3pt,linecolor=black]{-}{naa}{nb}
\ncline[linewidth=3pt,linecolor=black]{->}{nc}{nccc}
\ncline[linewidth=3pt,linecolor=black]{-}{ncc}{nd}\Aput{$\mathbf\nu$}
 
\rput(.8,.25){%
       \psellipse[linewidth=2pt,fillstyle=solid,fillcolor=lightgreen](2.5,0.2)(1.3,.8)
       \psellipse[fillstyle=solid,fillcolor=lightred](2.5,0.2)(.48,.28)
\psdots*[dotstyle=+,linewidth=5pt,dotscale=.3](2.5,0.2)
\psdots*[dotstyle=|,dotangle=90,linewidth=3pt,dotscale=.4]
(1.6,0.2)(3.4,0.2)(2.5,0.7)(2.5,-0.3)(3.0,-0.2)(3.0,0.6)(2.0,-0.2)(2.0,0.6)
}%

\bigskip
\bigskip

In the static limit, the reduced Dirac form factor $V^D(q^2)$ and the
neutrino anapole form factor $A^D(q^2)$ are related to the vector and
axial vector charge radius $\left< r^2_V \right>$ and $\left< r^2_A
\right>$ through:
\begin{equation}
\label{radii}
\left< r^2_V \right> = -6\, V^D(0);  \hspace{1truecm} 
\left< r^2_A \right> = -6\, A^D(0).
\end{equation}
In the following we will refer to these form factors as the vector and axial
vector charge radius also when $q^2\neq 0$.  

A long standing controversy about the possibility of consistently
defining a gauge invariant, physical, and process independent neutrino
charge radius \cite{controversy} has been recently settled
\cite{Bernabeu:2000hf,Bernabeu:2002nw,Cabral-Rosetti:2002qx,Bernabeu:2002pd}.\footnote{
In the 2002 Review of Particle Physics \cite{Hagiwara:pw} the limits
on the neutrino charge radius are reported as limits on ``nonstandard
contributions to neutrino scattering". The corresponding section
begins with the statement: ``We report limits on the
so-called neutrino charge radius squared in this section. This
quantity is not an observable physical quantity, and this is reflected
in the fact that it is gauge dependent.''}  The controversy was
related to the general problem of defining improved one-loop Born
amplitudes in $SU(2)\times U(1)$ for processes like e.g.
$e^+e^-\to f\bar f$.  Let us consider schematically the amplitude for
this process in QED.  We can include the leading one-loop corrections
while retaining a Born-like form simply by redefining the electromagnetic
coupling:
\begin{eqnarray}
{\cal M}^0_{QED}\hspace{-3mm} &=&\hspace{-3mm}e^2_0 \frac{J^e_Q\times
  J^f_Q}{q^2}\ \ \longrightarrow\ \  
{\cal M}_{QED}^1= e^2 \frac{J^e_Q \times J^f_Q}{q^2}, \nonumber \\
\frac{1}{e^2}\hspace{-3mm}&=&\hspace{-3mm} 
\frac{1}{e^2_0} - \Pi'_\gamma(q^2)\,,  
\end{eqnarray}
where $\Pi'_\gamma(q^2) $ is the reduced photon self energy.  In this
approach it is consistent to neglect the additional corrections due to
the photons box diagrams, since they are separately gauge invariant.
The same is not possible in the SM. A factorized Born-like expression
(with suitably renormalized couplings) for the one-loop neutral
current amplitude of the form:
\begin{equation}\label{factorized} 
{\cal M}_{NC}^1 =  e^2\ \frac{ J_Q^e \times J_Q^f}{q^2}  +
\left(\frac{e^2}{s^2_w c^2_w}\right) 
\frac{ J_Z^e \times J_Z^f}
{q^2 - M_Z^2 -iM_Z\Gamma_Z} 
\end{equation}
(with $f\neq e\,,\nu_e$) in general is not gauge invariant !

A sketchy argument to explain the problem goes as follows: if one
tries to define for example an improved effective $e^+e^-Z$ coupling
that includes the one-loop vertex corrections (fig. 1a) one finds that
the `improper vertex' corresponding to the mixed $\gamma$-$Z$ self
energy (fig. 1b) should also be included in order to obtain a finite
expression.  Still, the result is gauge dependent, and gauge
invariance can be recovered only by including also the $W$ box
diagrams (fig. 1c).  One can easily convince himself
\begin{figure}[h]
  \includegraphics[height=.145\textheight]{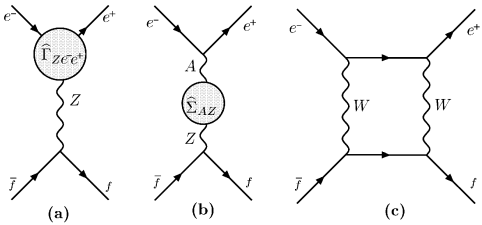}
  \caption{
    The vertex (a), mixed $A$-$Z$ self-energy (b) and $W$ box diagrams (c)
    that contribute to construct a gauge invariant 1-loop amplitude.}
\label{Fig1}
\end{figure}

\noindent
that things work in this way, by considering the cut of the diagrams
sketched in fig. 1 that involve $W$ bosons, as depicted in fig. 2. Each
single diagram obtained by cutting the vertex, the mixed self-energy
and the $W$ box represents only a partial contribution to the full
amplitude for the process $e^+ e^- \to W^+ W^-$, and of course a
single partial contribution has no well defined physical meaning: only
the full amplitude is a physical quantity.  On the other hand, the $W$
box diagrams needed to recover gauge invariance connect initial state
fermions to the final states, and therefore depend on the specific
process. This indicates that at one-loop, the amplitude cannot be
straightforwardly written in the factorized form (\ref{factorized}),
and that the definition of an improved and process independent
$e^+e^-Z$ vertex is indeed problematic.
\begin{figure}[h]
  \includegraphics[height=.16\textheight]{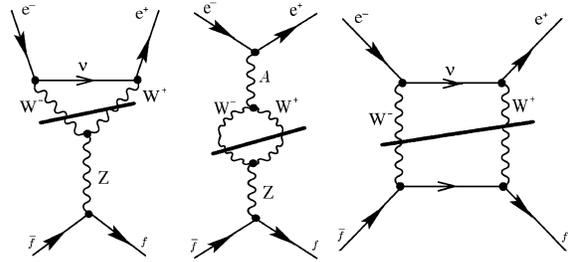}
  \caption{
A subset of diagrams contributing at 1-loop to 
$e^+e^- \to f\bar f$, cut to show the relation with 
$e^+e^- \to W^+W^-$.} 
\label{Fig2}
\end{figure}
This problem is even more acute when one tries to define the charge
radius of a neutrino as a physical, process independent property,
intrinsic to this particle. This is due to the absence at tree level
of a neutrino-photon coupling, so that there is no well defined
quantity even at the lowest order.

Consistent definitions of  one-loop gauge invariant vertexes,
self-energies and box diagrams in the electroweak sector of the SM
were first obtained by Deagrassi and Sirlin \cite{Degrassi:1992ue}
through the systematic application of a procedure 
previously developed in QCD, the so called pinch technique
\cite{Cornwall:1981zr}.  More recently, the same technique was used to
construct a neutrino charge radius at one-loop, which is independent of
the gauge fixing parameter, of the gauge fixing scheme, and of the
particular scattering process \cite{Bernabeu:2000hf}.  That such a
construction is possible can be intuitively understood by observing
that for neutrino scattering off right handed polarized fermions, the
$W$ box diagrams are absent to begin with, and thus no ambiguity can
arise \cite{Bernabeu:2000hf}. This  suggested a way to
derive a unique decomposition of loop contributions that separately
respect gauge invariance, and from which a process independent charge
radius could be defined as an intrinsic property of the neutrino.
Furthermore, in \cite{Bernabeu:2002nw,Bernabeu:2002pd} it was argued
that the so-defined charge radius is a true physical observable, in
the sense that its value could be extracted, at least in principle,
from experiments.

Coming back to the neutrino form factors appearing in
(\ref{JmuDirac}), not all of them survive when the neutrinos are
Majorana particles ($\nu^M$).  In the non-diagonal case
($\nu^M_j\neq\nu^M_i$) and in the limit of CP invariance the
electromagnetic interaction is described by just two form factors
\cite{Nieves:1981zt}.  If the initial and final Majorana neutrinos
involved in the process have the same CP parity, only $E_{ji}^M(q^2)$
and $A_{ji}^M(q^2)$ are non vanishing, while if the CP parity is
opposite, the electromagnetic interaction is described by
$M_{ji}^M(q^2)$ and $V_{ji}^M(q^2)$.  Finally, in the Majorana
diagonal case $\nu^M_j=\nu^M_i$ the only surviving form factor is the
anapole moment $A^M(q^2)$.  As discussed in \cite{Kayser:1982br}, this
last result can be inferred from the requirement that the two
identical fermions final state in $\gamma \to \nu^M \bar{\nu}^M$ be
antisymmetric, and it holds regardless of the assumption of CP
invariance.

In the SM the neutrino electromagnetic form factors have extremely
small values \cite{Dubovik:1996gx}.  Due to the left-handed nature of
the weak interactions, the numerical value of the vector and axial
vector charge radius coincide, and for the different $\nu_e$, $\nu_\mu
$ and $\nu_\tau $ flavors they fall within the range $\langle
r^2_{V,A} \rangle \approx (1 - 4)\times 10^{-33} {\rm cm^2}$
\cite{Bernabeu:2000hf}\footnote{These values correspond to the $q^2=0$
limit, and decrease with increasing energies with a logarithmic
behavior.}.

However, it could well be that, due to new physics, the
electromagnetic interactions of neutrinos are enhanced with respect to
the SM expectations.  In general, the strongest limits on anomalous
form factors come from astrophysical and cosmological considerations.
For example the neutrino magnetic moments can be constrained from
consideration of stellar energy losses through plasma photon decay
$\gamma \to \nu\bar \nu $ \cite{Raffelt:gv}, from the non-observation
of anomalous energy loss in the Supernova 1987A neutrino burst as
would have resulted from the rapid emission of superweakly interacting
right handed neutrinos \cite{Raffelt:gv}, or from Big Bang
nucleosynthesis arguments.  In the last case constraints are obtained
by requiring that spin flipping Dirac magnetic moment interactions
should be weak enough not to populate the right handed neutrinos
degrees of freedom at the time of the freeze out of the
neutron-to-proton ratio \cite{Dolgov:2002wy}.

Since the charge radii do not couple neutrinos to on-shell photons,
the corresponding interactions are not relevant for stellar evolution
arguments.  However, in the Dirac case, right handed neutrinos can
still be produced through e.g. $e^+e^-\to \nu_R\bar \nu_R$, and
therefore constraints from the Supernova 1987A as well as from
nucleosynthesis do apply.  They yield respectively $|\left< r^2
\right>| \lsim 2\times 10^{-33}{\rm cm}^2$ \cite{Grifols:1989vi} and
$|\left< r^2 \right>| \lsim 7\times 10^{-33}{\rm cm}^2$
\cite{Grifols:1986ed}\footnote{In the SM with right handed neutrinos
the $\nu_R$ cannot be produced through the charge radius couplings,
since the vector and axial vector contributions exactly
cancel. Therefore, the quoted limits implicitly assume that, because
of new physics contributions, one of the two form factors dominates
and no cancellations occur.}.

However, if neutrinos are Majorana particles, they do not have light
right-handed partners, and the previous constraints do not apply.  In
this case, in particular for $\nu_\tau$, a large anapole moment
resulting in an interaction even stronger than electroweak could be
allowed.  In the early Universe such an interaction could keep
$\nu_\tau$ in thermal equilibrium long enough to experience a
substantial reheating from the annihilation process $e^+e^- \to
\nu_\tau\bar \nu_\tau$, and in turn this could affect the Universe
expansion and change the abundance of primordial $^4He$.  In section 2
we will argue that this is not so, and that even an interaction one
order of magnitude stronger than electroweak would hardly affect
Helium abundance at the observable level.  We can conclude that
constraints on the Majorana neutrino axial charge radius can be
obtained only from terrestrial experiments.

The present laboratory limits for the electron neutrino are
$-5.5\times 10^{-32} \leq \langle r^2_A (\nu_e) \rangle \leq 9.8\times
10^{-32} {\rm cm}^2$ \cite{Allen:qe}\footnote{These limits are twice
the values published in \cite{Allen:qe} since we are using a
convention for $\langle r^2_{V,A}\rangle$ that differs for a factor of
2.}.  In the Dirac case these limits apply to the sum $\langle r^2_V
\rangle + \langle r^2_A\rangle$ as well.  Limits for $\nu_\mu$ have
been derived from neutrino scattering experiments
\cite{Vilain:1994hm,Ahrens:fp}.  They are about one order of magnitude
stronger than for $\nu_e$, and will be discussed in section 4.  Due to
the fact that intense $\nu_\tau$ beams are not available in
laboratories, to date no direct limits on $\langle r^2_A(\nu_\tau)
\rangle $ have been reported by experimental collaborations.  However,
under the assumption that a significant fraction of the solar
neutrinos converts into $\nu_\tau$, the limit $|\langle r^2_A (\nu_\tau)
\rangle | \lsim 2\times 10^{-31} {\rm cm}^2 $ was derived from the
SNO and Super-Kamiokande data \cite{Joshipura:2001ee}.  A limit on the
$\nu_\tau$ vector charge radius (Dirac case) was obtained from TRISTAN
data on $e^+ e^- \to \nu \bar \nu \gamma$ single photon production
\cite{Tanimoto:2000am}. As is discussed in section 3, the same data can
be used to set limits also on the anapole moment of a Majorana $\nu_\tau$.

In the next section we will briefly analyze the possibility of
deriving constraints on the axial charge radius of Majorana neutrinos
from nucleosynthesis.  In section 3 we will study the bounds on the
tau neutrino charge radius implied by the TRISTAN and LEP experimental
results.  In section 4 we will discuss the constraints on the muon
neutrino charge radius from the NuTeV, CHARM-II, CCFR and the BNL E734
experiments.  They result in the following 90\% c.l.  limits:
\begin{eqnarray}
\hskip-4mm  -8.2 \times 10^{-32}{\rm cm}^2 
\leq \hskip-3mm  &\left< r^2_A(\nu_\tau) \right>&\hskip-2mm   \leq
9.9 \times 10^{-32} {\rm cm}^2,  \label{limittau} \\
\hskip-4mm  -5.2 \times 10^{-33}  {\rm cm}^2 
\leq  \hskip-3mm  &\left< r^2_A(\nu_\mu) \right>& \hskip-2mm  \leq
 6.8 \times 10^{-33}  {\rm cm}^2. \label{limitmu}
\end{eqnarray}
For $\langle r^2_A (\nu_e)\rangle $ we could not find new experimental
results that would imply better constraints than the existing ones
\cite{Allen:qe}. We just mention that the Bugey nuclear reactor data
from the detector module closest to the neutrino source (15 meters)
\cite{Declais:1994su} should imply independent limits of the same
order of magnitude than the existing ones.
%
%
 \section{Nucleosynthesis}
 \label{sec:nucleo}
 
 In this section we study the impact on the primordial Helium
 abundance $Y$ of an axial charge radius large enough to keep a
 Majorana $\nu_\tau$ in thermal contact with the plasma down to
 temperatures $T < 1\,$MeV.  In this case the neutrinos would get
 reheated by $e^+ e^-$ annihilation, and this would affect the
 Universe expansion rate. To give an example, assuming that one
 neutrino species keeps in thermal equilibrium until $e^+ e^-$
 annihilation is completed ($T\ll m_e$) has the same effect on the
 expansion as $\Delta \nu=1-(4/11)^{4/3}\simeq 0.74$ additional
 neutrinos.
 
 The amount of Helium produced in the early Universe is determined by
 the value of the neutron to proton ratio $n/p$ at the time when the
 $ne^+ \leftrightarrow p\bar \nu$ and $n\nu \leftrightarrow pe^-$
 electroweak reactions freeze out.  This occurs approximately at a
 temperature $T_{fo}\approx 0.7\,$MeV \cite{Dicus:bz,Kolb:vq}.  Apart
 for the effect of neutron decay, virtually all the surviving neutrons
 end up in $^4He$ nuclei.  Assuming no anomalous contributions to the
 electron neutrino reactions, the freeze out temperature can only be
 affected by changes in the Universe expansion rate, that is
 controlled by the number of relativistic degrees of freedom and by
 their temperature.  If $\nu_\tau$ have only standard interactions, at
 the time of the freeze out they are completely decoupled from the
 thermal plasma. However, an anomalous contribution to the process
 $e^+e^-\leftrightarrow \nu_\tau \bar \nu_\tau$ would allow them to
 share part of the entropy released in $e^+e^-$ annihilation. The
 maximum effect is achieved assuming that the new interaction is able
 to keep the $\nu_\tau$ thermalized down to $T_{fo}$.  The required
 strength of the new interaction can be estimated by equating the rate
 for an anomalously fast $e^+e^-\leftrightarrow \nu_\tau \bar
 \nu_\tau$ process $\Gamma_{\nu_\tau}=\left<\sigma v\right> n_e$ to
 the Universe expansion rate $\Gamma_{U}= \left(8 \pi \rho/3
 m^2_P\right)^{1/2}$.  In the previous formulas $\left<\sigma
 v\right>$ is the thermally averaged cross section times relative
 velocity, $\,n_e\approx 0.365\, T^3$ is the number density of
 electrons, $\rho\approx 1.66\, g_*^{1/2}\, (T^2/m_P) $ is the
 Universe energy density with $g_*\approx 10.75$ the number of
 relativistic degrees of freedom, and $m_P$ is the Plank mass.  The
 thermally averaged cross section can be written as $\left<\sigma
 v\right>\simeq \kappa\, G^2_{\nu_\tau} T^2 $ where
 $G_{\nu_\tau}\approx (2\pi^2 \alpha/3) \left< r^2_A \right>$
 parametrizes the strength of the interaction that we assume sensibly
 larger than the Fermi constant $G_F$, and $\kappa\approx 0.2$ has
 been introduced to allow direct comparison with the SM rate
 $\left<\sigma v\right>^{SM}\simeq 0.2\, G^2_F\,
 T^2\,$\cite{Dicus:bz}.  By setting $\Gamma_{\nu_\tau}=\Gamma_{U}$ at
 $T=T_{fo}$, we obtain $G_{\nu_\tau}\approx 13 \times
 10^{-5}\,$GeV$^{-2}$.  Therefore, to keep the $\nu_\tau$ thermalized
 until the ratio $n/p$ freezes out, an interaction about ten times
 stronger than electroweak is needed.  However, even in the presence
 of such a large interaction, Helium abundance would only be mildly
 affected.  This is because at $T\approx 0.7\,$MeV $e^+e^-$
 annihilation is still not very efficient, and the photon temperature
 is only slightly above the temperature of the thermally decoupled
 neutrinos: $(T_\gamma-T_\nu)/T_\gamma\approx1.5\% $ \cite{Dicus:bz}.
 This induces a change in the primordial Helium abundance $\Delta
 Y\approx + 0.04\, (\Delta T_{\nu_\tau}/T_\nu)$ which is below one
 part in one thousand.  This effect could possibly be at the level of
 the present theoretical precision \cite{Lopez:1998vk}; however, it is
 far below the present observational accuracy, for which the errors
 are of the order of one percent \cite{Dolgov:2002cv}.

%
%
\section{Limits on $\nu_{\tau}$ vector and axial vector charge radius}
\label{nutau}

Limits on $\ncr$ and $\acr$ for $\nu_{\tau}$ can be set using
experimental data on single photon production through the process $e^+
e^- \to \bar \nu \nu \gamma$.  A large set of data from TRISTAN and
from LEP, spanning the energy range from 58 GeV up to 207 GeV, was
analyzed in \cite{Hirsch:2002uv}.  Given that the form factors run
with the energy, separate results were presented for data collected in
different energy ranges: below $Z$ resonance (TRISTAN), between $Z$
resonance and the threshold for $W^+W^-$ production (LEP-1.5), and
finally for the data above $W^+W^-$ production (LEP-2).  Due to the
much larger statistics collected at high energy, a combined fit of all
the data does not give any sizable improvement with respect to the
LEP-2 limits, that therefore represent the strongest bounds.

The SM cross section for the process $e^+e^- \to \nu{\bar \nu} \gamma$ is
given by \cite{Gaemers:fe}
\begin{eqnarray}
\label{ddsigSM}
\frac{d\sigma_{\nu\nu\gamma}}{dx\,dy} &=& \frac{2\alpha/\pi
}{x(1-y^2)}\left[\left(1-\frac{x}{2}\right)^2+\frac{x^2 y^2}{4}
\right]\times \nonumber \\ && \hspace{-8mm} \Big\{N_\nu\,
\sigma_s(s',g_V,g_A) + \sigma_{st}(s') + \sigma_t(s')\Big\}
\end{eqnarray}
where $\sigma_s$ corresponds to the lowest order $s$ channel $Z$ boson
exchange with $N_\nu=3$ the number of neutrinos that couple to the $Z$
boson.  For later convenience in $\sigma_s$ we have explicitly shown
the dependence on the electron couplings $g_V=-1/2+2 \sin^2\theta_W$
and $g_A=-1/2$, where $\theta_W$ is the weak mixing angle.  The
additional two terms $\sigma_{st}$ and $\sigma_t$ in (\ref{ddsigSM})
correspond respectively to $Z$-$W$ interference and to $t$ channel $W$
boson exchange in $\nu_e$ production.  The kinematic variables are the
scaled photon momentum $x=E_{\gamma}/E_{\rm beam}$ with $E_{\rm beam}
= \sqrt{s}/2$, the reduced center of mass energy $s'=s(1-x)$, and the
cosine of the angle between the photon momentum and the incident beam
direction $y=\cos\theta_{\gamma}$.  The expressions for the lowest
order cross sections appearing in (\ref{ddsigSM}) read
\begin{eqnarray}
\sigma_s(s) &=& \frac{s\,G_F^2 }{6\pi}\> \frac{\frac{1}{2}\, (g_V^2+g_A^2)\, M^4_Z}
{\left(M^2_Z-s\right)^2+ M_Z^2\Gamma_Z^2}\,, \label{sigs} \\ [5pt]
\sigma_{st}(s) &=& \frac{s\,G_F^2 }{6\pi}\>  \frac{(g_V+g_A)\,(M^2_Z-s)\,M^2_Z}
{\left(M^2_Z-s\right)^2+ M_Z^2\Gamma_Z^2}\,, \label{sigst} \\ [5pt]
\sigma_{t}(s) &=& \frac{s\,G_F^2 }{6\pi}\,,    \label{sigt}
\end{eqnarray}
where $G_F$ is the Fermi constant, $\alpha$ the fine structure
constant, $M_Z$ and $\Gamma_Z$ the mass and width of the $Z$ boson.
Few comments are in order. Eq.  (\ref{ddsigSM}) was first derived in
\cite{Gaemers:fe}. It holds at relatively low energies where $W$
exchange in the $t$ channel can be legitimately approximated as a
contact interaction.  This amounts to neglect the momentum transfer in
the $W$ propagator, and to drop the $W$-$\gamma$ interaction, so that
the photons are emitted only from the electron lines.  While this
approximation is sufficiently good at TRISTAN energies, to analyze the
LEP data collected above $Z$ resonance some improvements have to be
introduced.  We will use an improved approximation where finite
distance effects are taken into account in the $W$ propagator, however
we will still work in the limit of vanishing $W$-$\gamma$
interactions, since the corresponding effects are of higher order in a
leading log approximation \cite{Bardin:2001vt} and for our scopes can
be safely neglected.  Finite distance $W$ exchange effects can be
taken into account in the previous expressions by replacing
$\sigma_{st}(s)$ and $\sigma_{t}(s)$ by
%
\begin{equation}
\label{finitedist}
\sigma_{st}(s) \cdot F_{st} 
\left(\frac{s}{M^2_W}\right), \quad\  {\rm and} \quad\ 
\sigma_{t}(s) \cdot F_{t}
\left(\frac{s}{M^2_W}\right)\,,    
\end{equation}
respectively, where $M_W$ is the $W$ boson mass, and 
\begin{eqnarray}
\hspace{-3mm}
F_{st}(z) \hspace{-2mm} &=& \hspace{-2mm}\frac{3}{z^3} \Big[(1+z)^2\log(1+z) -
z\,\Big(1+\frac{3}{2}\,z\Big) \Big]\,, \label{fst}  \\  
\hspace{-3mm}
F_{t}(z) \hspace{-2mm} &=&  \hspace{-2mm}
\frac{3}{z^3} \Big[-2\,(1+z)\,\log(1+z) + z\,(2+z) \Big]\,.   \label{ft}
\end{eqnarray}
The contact interaction approximation is recovered in the limit $z\to
0$ for which $F_{st,t}(z) \to 1$.

An anomalous interaction due to non-vanishing $\nu_{\tau}$ axial and
axial vector charge radii can be directly included in (\ref{ddsigSM})
by redefining the $Z$ boson exchange term in the following way:
\begin{eqnarray}
&& N_\nu\, \sigma_s(s',g_V,g_A) \ \longrightarrow\nonumber \\
&& (N_\nu-1)\>\sigma_s(s',g_V,g_A)\, +\, 
  \sigma_s(s', g_V^*(s'),g_A) \hspace{1truecm} \label{newPhys}
\end{eqnarray}
where
\begin{eqnarray}
 g_V^*(s') &=& g_V - \left[1-\frac{s'}{M^2_Z}\right]\, \delta\,, \label{Deftildev}\\ 
\delta &=& \frac{\sqrt{2}\pi\alpha}{3 G_F}\left[\ncr +\acr\right]\,. \label{DefDelta}
\end{eqnarray}
The substitution $g_V\to g_V^*$ in (\ref{newPhys}) takes into account
the new photon exchange diagram for production of left-handed
$\nu_\tau$.  In the Dirac case, $s$-channel production of right handed
$\nu_\tau$ through photon exchange must also be taken into account.
This yields a new contribution that adds incoherently to the cross
section, and that can be included by adding inside the brackets in
(\ref{ddsigSM}) the term
\begin{eqnarray} 
\sigma_R\,(s') &=&  \frac{s'\,G_F^2 }{6\pi}\,  (\delta')^2\,,\label{nuR} \\ 
\delta' &=& \frac{\sqrt{2}\pi\alpha}{3 G_F}\left[\ncr -\acr\right].\label{DefDeltaP}
\end{eqnarray}
In the SM $\ncr = \acr$ and therefore there is no production of
$\nu_R$ through these couplings.  For a Majorana neutrino $\delta'=0$
and $\ncr = 0$, and thus the limits on anomalous contributions to the
process $e^+e^- \to \nu {\bar \nu} \gamma$ translate into direct
constraints on the axial charge radius $\acrt$.

\subsection{Limits from TRISTAN}

The three TRISTAN experiments AMY \cite{Sugimoto:1995bf}, TOPAZ
\cite{Abe:fq} and VENUS \cite{Hosoda:1994bd} have searched for single
photon production in $e^+e^-$ annihilation at a c.m. energy of
approximately $\sqrt{s}=58\,$GeV.  Anomalous contributions to the
cross section for $e^+e^- \to \nu{\bar \nu} \gamma$ would have been
signaled by an excess of events.  Limits on a Dirac $\nu_\tau$ charge
radius from the TRISTAN data were derived in \cite{Tanimoto:2000am}.
The analysis was extended in \cite{Hirsch:2002uv} to the include the
case of a Majorana neutrino with an anapole moment.  The details of
TRISTAN results can be found in table 1 of ref. \cite{Hirsch:2002uv}.
For a Majorana $\nu_\tau$ ($\delta'=0$ and $\ncr = 0$) the data imply
the following 90 \% c.l.  limits:
\begin{equation}
-3.7 \times 10^{-31} \hskip1pt {\rm cm}^2 \hskip-1pt \leq \acrt \leq\hskip-1pt
 3.1 \times 10^{-31} \hskip1pt  {\rm cm}^2.  
\label{TristanMaj}
\end{equation}
For the Dirac case, the associated production of right-handed states
through $\sigma_R$ in (\ref{nuR}) allows us to constrain independently
the vector and axial vector charge radius.  The 90 \% c.l.  limits on
$ \vacrt \equiv \ncrt + \acrt$ are:
\begin{equation}  \label{vacrtTRIS} 
-2.1 \times 10^{-31}\hskip1pt  {\rm cm}^2 \hskip-1pt \leq \vacrt \leq
\hskip-1pt 1.8 \times 10^{-31} \hskip1pt {\rm cm}^2.               
\end{equation}
As we have already mentioned, strictly speaking the constraints just
derived cannot be directly compared with the LEP constraints analyzed
below, since the two experiments are proving neutrino form factors at
different energy scales.  Of course, since our limits constrain
essentially only physics beyond the SM, it is not possible to make a
sound guess of the form of the scaling of the form factors with the
energy, which is determined by the details of the underlying new
physics.  However, if we assume a logarithmic reduction of the form
factors with increasing energy as is the case in the SM, than we would
expect a moderate reduction of about $\approx 0.65$ when scaling from
TRISTAN to LEP-1.5 energies, and an additional reduction of about
$\approx 0.75$ from LEP-1.5 up to LEP-2 measurements at 200 GeV.

\subsection{Limits from LEP}

\nocite{Buskulic:1996hw,Adam:1996am,Alexander:1996kp,Ackerstaff:1997ze,Abbiendi:1998yu}
Limits on $\ncr$ and $\acr$ were derived from the observation of
single photon production at LEP in a completely similar way
\cite{Hirsch:2002uv}.  Contrary to magnetic moment interactions that
get enhanced at low energies with respect to electroweak interactions,
the interaction corresponding to a charge radius scale with energy
roughly in the same way than the electroweak interactions, and
therefore searches for possible effects at high energy are not in
disadvantage with respect to low energy experiments.  It is for this
reason that LEP data above the $Z$ resonance are able to set the best
constraints on the vector and axial vector charge radius for the
$\tau$ neutrino.

All LEP experiments have published high statistics data for the
process $e^+e^-\to \nu\bar\nu \gamma$ for c.m. energies close to the
$Z$-pole; however, due to the dominance of resonant $Z$ boson
exchange, these data are not useful to constrain anomalous neutrino
couplings to $s$-channel off-shell photons.  Therefore, only
off-resonance data collected above $Z$ resonance in the in the energy
range 130 GeV -- 207 GeV, were used in the analysis in
\cite{Hirsch:2002uv}.  The data were divided into two sets: LEP-1.5
data collected below $W^+W^-$ production threshold (table 2 of
ref. \cite{Hirsch:2002uv}) and LEP-2 data collected above $W^+W^-$
threshold up to 207 GeV (table 3 of ref. \cite{Hirsch:2002uv}).

\smallskip

{\bf LEP-1.5:} \ 
The ALEPH \cite{Buskulic:1996hw}, DELPHI \cite{Adam:1996am} and OPAL
\cite{Alexander:1996kp,Ackerstaff:1997ze,Abbiendi:1998yu}
collaborations have published data for single photon production at
c.m. energies of 130 GeV and 136 GeV.  During the fall 1995 runs ALEPH
\cite{Buskulic:1996hw} and DELPHI \cite{Adam:1996am} accumulated about
6 pb$^{-1}$ of data for each experiment, observing respectively 40 and
37 events.  In the same runs OPAL
\cite{Alexander:1996kp,Ackerstaff:1997ze} collected a little less than
5 pb$^{-1}$ observing 53 events.  In addition, OPAL published data
also for the 1997 runs (at the same energies) \cite{Abbiendi:1998yu}
collecting an integrated luminosity of 5.7 pb$^{-1}$ and observing 60
events.  With a total integrated luminosity of about 28 pb$^{-1}$
LEP-1.5 implies the following 90 \% c.l.  limits:
\begin{equation}
-5.9 \times 10^{-31}\hskip1pt  {\rm cm}^2 \hskip-1pt \leq \acrt \leq
\hskip-1pt 6.6 \times 10^{-31} \hskip1pt {\rm cm}^2 
\label{LEP-1.5Maj}
\end{equation}
for the axial vector charge radius of a Majorana $\nu_\tau $, and 
\begin{equation}   \label{vacrtLEP-1.5}   
 -3.5 \times 10^{-31}\hskip1pt  {\rm cm}^2 \hskip-1pt \leq \vacrt \leq
 \hskip-1pt 3.7 \times 10^{-31} \hskip1pt {\rm cm}^2             
 \end{equation}
for the Dirac case.  In spite of the much larger statistics, the
limits from LEP-1.5 (\ref{LEP-1.5Maj}) and (\ref{vacrtLEP-1.5}) are
roughly a factor of two worse than the limits from TRISTAN in
(\ref{TristanMaj}) and (\ref{vacrtTRIS}). The main reason for this is
that at LEP-1.5 energies initial state radiation tends to bring the
effective c.m. energy $s'$ of the collision close to the $Z$
resonance, thus enhancing $Z$ exchange with respect to the new photon
exchange diagram.
%
%

\smallskip

{\bf LEP-2:} \ 
Above the threshold for $W^+W^-$ production the four LEP experiments
reported results for 24 different data points, corresponding
altogether to about 1.6 nb$^{-1}$ of data (see table 3 of ref.
\cite{Hirsch:2002uv}).

ALEPH \cite{Barate:1997ue,Barate:1998ci,Heister:2002ut} published data
for ten different c.m. energies, ranging from 161 GeV up to 209 GeV.
DELPHI \cite{Abreu:2000vk} published data collected at 183 GeV and 189
GeV, and gave separate results for the three major electromagnetic
calorimeters, the High density Projection Chamber (HPC) covering large
polar angles, the Forward ElectroMagnetic Calorimeter (FEMC) covering
small polar angles, and the Small angle TIle Calorimeter (STIC) that
covers the very forward regions.  In three papers
\cite{Acciarri:1997dq,Acciarri:1998hb,Acciarri:1999kp} L3 reported the
results obtained at 161 GeV, 172 GeV, 183 GeV and 189 GeV.  Finally,
OPAL published data for four different c.m. energies
\cite{Alexander:1996kp,Ackerstaff:1997ze,Abbiendi:1998yu,Abbiendi:2000hh}.
The 90 \% c.l. limits implied by LEP-2 data read
\begin{equation}
-8.2 \times 10^{-32}\hskip1pt  {\rm cm}^2 \hskip-1pt \leq \acrt \leq
\hskip-1pt 9.9 \times 10^{-32} \hskip1pt {\rm cm}^2 
\label{LEP-2Maj}
\end{equation}
%
%
for the Majorana case, and   
\begin{equation} \label{vacrtLEP-2}  
-5.6 \times 10^{-32}\hskip1pt  {\rm cm}^2 \hskip-1pt \leq \vacrt \leq
\hskip-1pt 6.2 \times 10^{-32} \hskip1pt {\rm cm}^2     
\end{equation}
for a Dirac $\nu_\tau$. 

These limits are about a factor of four stronger than the limits
derived in \cite{Joshipura:2001ee} from the SNO and Super-Kamiokande
observations and than the limits obtained in \cite{Tanimoto:2000am}
from just the TRISTAN data.

It is worth mentioning that the limits on $\ncrt$ and $\acrt$ are
almost uncorrelated between them (see figure 3 of
ref. \cite{Hirsch:2002uv}).  The possibility of bounding
simultaneously the vector and axial vector charge radii stems from the
fact that in $e^+ e^-$ annihilation also the right-handed neutrinos
can be produced, and they couple to the photon through a combination
of $\ncr$ and $\acr$ which is linearly independent with respect to the
one that couples the left-handed neutrinos.  In contrast, neutrino
scattering experiments do not involve the right handed neutrinos, and
therefore can only bound the combination $\ncr + \acr$.

%
%
Before concluding this section, we should mention that independent
limits could also be derived from the DONUT experiment, through an
analysis similar to the one presented in \cite{Schwienhorst:2001sj} 
that yielded limits on the $\nu_\tau$ magnetic moment.  We have
estimated that the constraints from DONUT would be at least one order
of magnitude worse than the limits obtained from LEP; however, it
should be remarked that these limits would be inferred directly from
the absence of anomalous interactions for a neutrino beam with an
identified $\nu_\tau$ component \cite{Kodama:2000mp}.
%
%

\section{Limits on $\nu_{\mu}$ vector and axial vector charge radius}

The NuTeV collaboration has recently published a value of $\swsq$
measured from the ratio of neutral current to charged current in deep
inelastic $\nu_{\mu}$-nucleon scattering \cite{Zeller:2001hh}.  Their
result reads
\begin{equation}
\sin^2\theta_W^{(\nu)} = 0.2277 \pm 0.0013 \pm 0.0009
\label{NuTeVRes}
\end{equation}
where the first error is statistical and the second error is
systematic.  In order to derive limits on neutrino electromagnetic
properties one should compare the results obtained in neutrino
experiments to a value of $\swsq$ determined from experiments that do
not involve neutrinos.  Currently, the most precise value of $\swsq$
from non-neutrino experiments comes from measurements at the $Z$-pole
and from direct measurements of the $W$-mass \cite{Hagiwara:pw}.  In
the numerical calculations in \cite{Hirsch:2002uv} we have used the
value for $\swsq$ obtained from a global fit to electroweak
measurements without neutrino-nucleon scattering data, as reported in
\cite{Zeller:2001hh,Zeller:2002dx}:
\begin{equation}
\swsq = 0.2227 \pm 0.00037\,.
\label{EWWGsw}
\end{equation}
The effect of a non-vanishing charge radius can be taken into account
through the replacement $g_V \to g_V - \delta$ in the formulas for
$\nu_\mu$-nucleon and $\nu_\mu$-electron scattering \cite{Vogel:iv},
where $\delta$ is given in (\ref{DefDelta}).  Since there are no
right-handed neutrinos involved, there is no effect proportional to
$\delta'$ and therefore only $\delta \propto \ncrm + \acrm$ can be
constrained.  Upper and lower limits can be directly derived by
comparing $\sin^2\theta_W^{(\nu)}$ with the quoted value of $\swsq$
from non-neutrino experiments.  Since the results for neutrino
experiments and the measurements at the $Z$-pole are not consistent at
the 1$\sigma$ level, in the following equations
(\ref{LimNuTeV})-(\ref{LimCharm}) we will (conservatively) combine the
errors by adding them linearly.\footnote{Except for the CCFR data,
which is consistent with the SM precision fits.}
%
\begin{figure}
 \hspace{-35mm}
 \vspace{0mm}
  \includegraphics[height=.187\textheight]{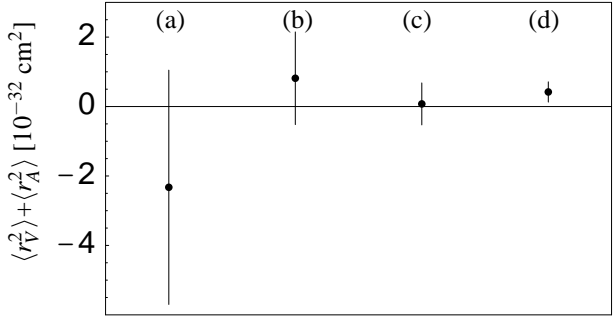}
 \hspace{-79mm}
\begin{rotate}{90}
$\phantom{aaaa}  \ncr$+$\acr$ [$10^{-32}$ cm$^2$]
\end{rotate}
\put(20,109){ \phantom{aaaaa} (a) \hskip12mm (b) \hskip11mm (c) \hskip12mm (d) }
  \caption{
90 \% c.l. limits on ($\ncr$+$\acr$) for the muon neutrino derived from (a)
  E734 at BNL \cite{Ahrens:fp}, (b) CHARM II \cite{Vilain:1994hm}, (c) CCFR
  experiment \cite{McFarland:1997wx} and (d) from the NuTeV result
  \cite{Zeller:2001hh}
}
\label{NuTevLim}
\end{figure}
%
%

From the NuTeV result (\ref{NuTeVRes}) we obtain the 90 \% c.l. upper
limit:
\begin{equation}
\vacrm \leq
\hskip2mm 7.1 \times 10^{-33} \hskip2mm {\rm cm}^2\,,  
\label{LimNuTeV}
\end{equation}
where $\vacrm\equiv \ncrm + \acrm$.  However, since (\ref{NuTeVRes})
hints to a non-vanishing value of $\delta$, no lower limit is obtained
from this measurement (see fig. \ref{NuTevLim}).  A reanalysis of the
E734 data on $\nu_\mu$-$e$ and $\bar \nu_\mu$-$e$ scattering
\cite{Ahrens:fp} yields the 90 \% c.l. limits:
\begin{equation}
-5.7 \times 10^{-32}  {\rm cm}^2 \hskip-1pt \leq \vacrm \leq \hskip-1pt
1.1 \times 10^{-32}  {\rm cm}^2.   
\label{LimAhr}
\end{equation}
Note that in ref. \cite{Ahrens:fp} the E734 collaboration is quoting a
lower limit about 3.6 times and an upper limit about 7.5 times tighter
than the ones given in (\ref{LimAhr}). This is because of various
reasons: first of all, as was pointed out in \cite{Allen:1990xn}, in
\cite{Ahrens:fp} an inconsistent value for $G_F$ was used that
resulted in bounds stronger by approximately a factor of
$\sqrt{2}$. In addition, the errors were combined quadratically,
which, due to the large negative trend in their data, resulted in a
much stronger upper bound on $\vacrm$ than the one quoted here.
Finally, our value of $\delta$ is defined through the shift $g_V \to
g_V - \delta$ of the SM vector coupling, consistently for example with
the notation of \cite{Vogel:iv}, while the convention used by the E734
Collaboration \cite{Ahrens:fp} as well as by CHARM II
\cite{Vilain:1994hm} define $\delta$ as a shift in $\swsq$. This
implies that our limits are larger for an additional factor of 2 with
respect to the results published by these two collaborations.

From the CHARM II neutrino-electron scattering data
\cite{Vilain:1994hm} we obtain at 90 \% c.l.:
\begin{equation}
-5.2 \times 10^{-33}  {\rm cm}^2 \hskip-1pt \leq \vacrm \leq
\hskip-1pt 2.2 \times 10^{-32} {\rm cm}^2.  
\label{LimCharm}
\end{equation}
These limits differ from the numbers published by the CHARM II
collaboration \cite{Vilain:1994hm} not only because of the mentioned
factor of 2 in the definition of $\delta$, but also because the
present value of $\swsq$ \cite{Hagiwara:pw} is smaller than the one
used in 1995 in the CHARM II analysis.

From the data published by the CCFR collaboration
\cite{McFarland:1997wx} one can deduce
\begin{equation}
-5.3 \times 10^{-33} {\rm cm}^2 \hskip-1pt \leq \vacrm \leq
\hskip-1pt 6.8 \times 10^{-33}  {\rm cm}^2.  
\label{LimCCFR}
\end{equation}
The four limits discussed above are represented in
fig. \ref{NuTevLim}, that makes apparent the level of precision of the
NuTeV result.  By combining the upper limit from CCFR (\ref{LimCCFR})
and the lower limit from CHARM II (\ref{LimCharm}) we finally obtain:
\begin{equation}
-5.2 \times 10^{-33}  {\rm cm}^2 \hskip-1pt \leq \vacrm \leq
\hskip-1pt 6.8 \times 10^{-33} {\rm cm}^2.  
\label{LimAll}
\end{equation}
It is well known that the NuTeV result shows a sizable deviation from
the SM predictions \cite{Zeller:2001hh}, and as a consequence it also
appears to be inconsistent (at the 90 \% c.l.) with $\delta = 0$. In
fact, strictly speaking their result
$\ncrm+\acrm = (4.20\pm 1.64)\times 10^{-33}$ cm$^2$ (1 $\sigma$ error)
could be interpreted as a measurement of $\ncrm+\acrm$, that becomes
consistent with zero only at approximately $2.5$ standard deviations.
However, while the quoted value is not in conflict with other
experimental limits, we believe that it would be not easy to construct
a model that could generate a neutrino charge radius of the required
size, without conflicting with other high precision electroweak
measurements.
%
%
%

\begin{theacknowledgments}
It is a pleasure to thank the Organizing Committee of the X Mexican
School of Particles and Fields for their kind invitation, for the
excellent organization and for the very pleasant atmosphere during the
conference.  This work was supported in part by COLCIENCIAS in
Colombia and by CSIC in Spain through a joint program for
international scientific cooperation.
\end{theacknowledgments}


\bibliographystyle{aipproc}   

\IfFileExists{\jobname.bbl}{}
 {\typeout{}
  \typeout{******************************************}
  \typeout{** Please run "bibtex \jobname" to optain}
  \typeout{** the bibliography and then re-run LaTeX}
  \typeout{** twice to fix the references!}
  \typeout{******************************************}
  \typeout{}
 }

\end{document}